\documentclass[pra,aps,twocolumn,english,superscriptaddress]{revtex4-1}
\usepackage[T1]{fontenc}
\usepackage[latin9]{inputenc}
\usepackage{geometry}
\usepackage{amsmath}
\usepackage{amsthm}
\usepackage{amssymb}
\usepackage{xcolor}
\usepackage[normalem]{ulem}
\def\oper{{\mathchoice{\rm 1\mskip-4mu l}{\rm 1\mskip-4mu l}
{\rm 1\mskip-4.5mu l}{\rm 1\mskip-5mu l}}}
\makeatletter

  \theoremstyle{plain}
  \newtheorem*{prop*}{\protect\propositionname}
 \theoremstyle{definition}
  
\newtheorem{thm}{Theorem}
\newtheorem{ex}{Example}
\newtheorem{pro}{Proposition}

\usepackage{braket}
\DeclareMathOperator{\Tr}{Tr}
\usepackage{amsmath}

\makeatother

\usepackage{babel}
  \providecommand{\examplename}{Example}
  \providecommand{\propositionname}{Proposition}

\newtheorem*{thm1}{Theorem 1}
\newtheorem*{thm2}{Theorem 2}

\begin{document}

\title{Optimal entanglement witnesses from limited local
measurements}

\author{Alberto Riccardi}
\email{alberto.riccardi01@universitadipavia.it}
\affiliation{Dipartimento di Fisica, University of Pavia, via Bassi 6, 27100 Pavia, Italy.}

\author{Dariusz Chru\'sci\'nski}
\affiliation{Institute of Physics, Faculty of Physics, Astronomy and Informatics,
Nicolaus Copernicus University, Grudziadzka 5/7, 87-100 Torun, Poland.}

\author{Chiara Macchiavello}
\affiliation{Dipartimento di Fisica, University of Pavia, via Bassi 6, 27100 Pavia, Italy.}
\affiliation{INFN Sezione di Pavia, via Bassi 6, I-27100, Pavia, Italy.}
\affiliation{CNR-INO, largo E. Fermi 6, I-50125, Firenze, Italy.}

\begin{abstract}
We address the problem of optimising entanglement witnesses when
a limited fixed set of local measurements can be performed on a bipartite
system, thus providing a procedure, feasible also for experiments, to detect entangled states using only the statistics of these local measurements. We completely characterize the class of entanglement witnesses
{of the form $W = P ^{\Gamma}$}, where $\Gamma$  denotes partial transposition,
that can be constructed from the measurements of the bipartite operators
$\sigma_{x}\otimes\sigma_{x},$
$\sigma_{y}\otimes\sigma_{y}$ and $\sigma_{z}\otimes\sigma_{z}$
in the case of two-qubit systems. In particular, we consider all possible extremal decomposable witnesses within the considered class that can be defined from this set of measurements.
Finally, we discuss possible extensions to higher dimension
bipartite systems when the set of available measurements is characterized
by the generalized Gell-Mann matrices. We provide several examples
of entanglement witnesses,
both decomposable and indecomposable, that can be constructed
with these limited resources.
\end{abstract}

\maketitle

\noindent {\em Introduction.} --- Entanglement is one of the unique aspects of quantum physics and nowadays
it is recognized as a pivotal resource in many quantum information
areas. Hence, testing whether a state of a composite system is separable
or entangled is a crucial task. A well-known procedure to accomplish
this task, without requiring full tomography of the state, uses
the notion of entanglement witnesses (EW) \cite{HHH,Terhal1,Terhal2}.
An operator $W$ that acts on a bipartite system living in $\mathcal{H}_A \otimes \mathcal{H}_B$ is an EW if and only if
 $W$ is a block-positive operator, i.e. $\langle \psi_A \otimes \psi_B | W | \psi_A \otimes \psi_B \rangle \geq 0$, but it is not positive, that is, it has at least one negative eigenvalue. Equivalently, $W$ is an EW iff i) $\Tr\left[W\rho_{sep}\right] \geq 0$
for all separable states $\rho_{sep}$, and  ii) there exists at least one entangled state $\rho_{e}$ such that $\Tr\left[W\rho_{e}\right]<0$.
Moreover, a state is entangled iff it is detected by some EW \cite{HHH}. Several types of EW have been defined and studied (see e.g. \cite{Guhne,Darek1} for the review). For example an EW is decomposable if $W = A + B^\Gamma$, where $A,B \geq 0$ and $B^\Gamma$ denotes a partial transposition. Witnesses which do not allow such decomposition are called
indecomposable. The former cannot detect positive partial transpose (PPT) entangled states, while
the latter can and therefore they can be useful to detect bound entangled states. In the case of two-qubit systems, since PPT entangled states do not exist, all entanglement witnesses are decomposable, hence they can be always represented as $W = A + B^\Gamma$.  Moreover, a witness $W$ is optimal if for any {$C \geq 0$} an operator  $W - C$ is no longer a witness \cite{LEW}.  Optimal EWs are the {\em best entanglement detectors}, that is,  $W$ is optimal if and only if there is no other witness that detects more quantum entangled states than $W$ does \cite{LEW}. Hence optimality has a direct operational meaning.
A witness which is not optimal may be {\em optimized} via a suitable optimization procedure \cite{LEW}. It is therefore clear that knowing all optimal EWs one is able to detect all entangled states. Among optimal witnesses there are extremal ones: an $W$ is extremal if for any block-positive {$D$} (not collinear with $W$) an operator  {$W - D$}  is no longer a witness. Extremal witnesses are fully characterised in the class of decomposable EWs: a decomposable EW is extremal if and only if $W = |\psi\rangle \langle \psi|^{\Gamma}$ for some entangled vector $\psi \in \mathcal{H}_A \otimes \mathcal{H}_B$. Characterization of extremal but indecomposable EWs is much more complicated and still open.\\

In this Letter we address the problem of defining the most general class of
EWs given a limited set of measurements. This work fits within the framework of
studying the experimental realization of quantum information tasks
when only a limited set of resources is available, as for example
the estimation of quantum channel capacities \cite{ChiaraMax1, ChiaraMax2}. More specifically,
we assume that a full tomography of the state of the system  is not
available, and our knowledge
is represented by the statistics of a set of local measurements $\mathcal{M}$.
The goal is to {process} the classical measurement outcomes in order to
find the most general classes of EWs that can be defined with this
set of local measurements.

\vspace{.1cm}

\noindent {\em Two qubits}. --- In this Letter we provide a complete characterization of EWs of the form $W = P ^{\Gamma}$ that can
be constructed for a two-qubit system by considering only the statistics
of $\mathcal{M}=\left\{ \sigma_{x}\otimes\sigma_{x},\sigma_{y}\otimes\sigma_{y},\sigma_{z}\otimes\sigma_{z}\right\}$.
Any 2-qubit entanglement witness $W$ can be represented as follows:
\begin{equation}\label{}
  W = \sum_{\mu,\nu=0}^3 T_{\mu\nu} \sigma_\mu \otimes \sigma_\nu ,
\end{equation}
with $\sigma_\mu \in \{\oper,\sigma_x,\sigma_y,\sigma_z\}$, and 16 real parameters $T_{\mu\nu}$.
Now, having an access to the limited resources $\mathcal{M}$ we consider witnesses with diagonal correlation tensor $T_{kl} = c_k \delta_{kl}$ $(k,l=1,2,3)$ only,  that is,

\begin{align} \label{WWW}
W = & \alpha\, \oper \otimes\oper +\sum_{k=x,y,z} \Big(  a_k \oper \otimes \sigma_k
 + b_k \sigma_k\otimes \oper \Big) \nonumber \\  & + \sum_{k=x,y,z} c_k  \sigma_k\otimes \sigma_k  ,
\end{align}
with real parameters $\alpha,a_k,b_k,c_k$. This reduces the number of independent parameters from 6 elements $T_{kl}=T_{lk}$ to 3 real parameters $c_k$. The above form is justified by the fact that the mean values of single qubit operators $\sigma_{k}\otimes\oper$ and
$\oper \otimes\sigma_{k}$  for $i=x,y,z$ can be derived by simply ignoring the statistics on one side.
Moreover, all the EWs within the class \eqref{WWW} can be also represented as $ W = P^{\Gamma} + Q$, with $P,Q \geq 0$. It should be stressed, that even if  $P$ and $Q$ do not belong to $\mathcal{M}$ it may happen that $ W = P^{\Gamma} + Q$ does \cite{G1,G2}. Hereafter, we consider only witnesses within the class \eqref{WWW} that are of the form
\begin{equation}
W = P^{\Gamma},
\label{WWW12}
\end{equation}
where $P$ contains only terms from  $\mathcal{M}$.
The {\em canonical} example of such a witness is provided by the flip operator
\begin{equation}\label{F-2}
  \mathbb{F} = \frac 12 \Big( \oper \otimes \oper + \sigma_x \otimes \sigma_x +  \sigma_y \otimes \sigma_y + \sigma_z \otimes \sigma_z \Big) .
\end{equation}
As is well known $\mathbb{F}$ witnesses entanglement within a class of Werner states:
\begin{equation}\label{werner}
  \rho_W = \frac{1}{2(2-f)} \Big( \oper \otimes \oper - f\, \mathbb{F}\Big) , \ \ -1 \leq f \leq 1 ,
\end{equation}
that is, $\rho_W$ is entangled iff ${\rm Tr}(\mathbb{F} \rho_W) < 0$ which is equivalent to $f > \frac 12$. Another well known example is an isotropic 2-qubit state
\begin{equation}\label{iso}
  \rho_{\rm iso} = \frac{1}{4} \oper \otimes \oper + r |\phi^+\rangle \langle \phi^+| ,
\end{equation}
where we denote by $\ket{\phi^{\pm}}$ and $\ket{\psi^{\pm}}$ the standard Bell states. The corresponding witness has the following form $W_{\rm iso} =  |\psi^-\rangle \langle \psi^-|^\Gamma$, and hence it belongs to our class

\begin{equation}\label{}
  W_{\rm iso} = \frac 12 \Big( \oper \otimes \oper - \sigma_1 \otimes \sigma_1 +  \sigma_2 \otimes \sigma_2 -  \sigma_3 \otimes \sigma_3 \Big) .
\end{equation}
Again $\rho_{\rm iso}$ is entangled iff ${\rm Tr}(\rho_{\rm iso} W_{\rm iso})<0$ which implies $r > \frac 13$.
To find EWs compatible with the above structure we find all state vectors $|\varphi\rangle$ such that $|\varphi\rangle \langle \varphi|$ has the structure (\ref{WWW}). It is evident that $|\varphi\rangle \langle \varphi|^\Gamma$ fits also the structure (\ref{WWW12}. Our main result states

\begin{thm} There are six 1-parameter families of rank-1 projectors $|\varphi\rangle\langle \varphi|$ of the form (\ref{WWW}):

\begin{alignat}{1}
\ket{\varphi_{1}}= a\ket{\phi^{+}}+b\ket{\phi^{-}}; &  \ket{\varphi_{2}}= a\ket{\psi^{+}}+b\ket{\psi^{-}};  \nonumber \\
\ket{\varphi_{3}}= a\ket{\phi^{+}}+b\ket{\psi^{+}}; & \ket{\varphi_{4}}= a\ket{\phi^{-}}+b\ket{\psi^{-}}; \nonumber \\
\ket{\varphi_{5}}= a\ket{\phi^{+}}+ib\ket{\psi^{-}}; & \ket{\varphi_{6}}= a\ket{\phi^{-}}+ib\ket{\psi^{+}}, \label{SS}
\end{alignat}
where $\ket{\phi^{\pm}}$ and $\ket{\psi^{\pm}}$ are the Bell states, and $a,b \in \mathbb{R}$ are such that $a^2+b^2=1$.\\
\end{thm}

Note that $|\varphi_k\rangle$ is entangled if and only if $a\neq\pm\frac{1}{\sqrt{2}}$. The proof of theorem $1$ is given in Supplementary Material. Interestingly,  the above sets of
states are the same as the ones {reported} in \cite{ChiaraMax1} in the context
of detection of quantum channel capacities with limited local measurements,
where the sets {were} derived by requiring orthogonality between states in order
to have bases of the Hilbert space. Here this condition is relaxed
but we arrive at the same result.

It is, therefore, clear that there are six families of extremal  EWs belonging to the class \eqref{WWW12}:
\begin{equation} \label{EWs}  W_k = |\varphi_k \rangle\langle \varphi_k|^\Gamma,
\end{equation}
for $k=1,\ldots,6$. For instance the extremal witness $W_1 = |\varphi_1\rangle\langle \varphi_1|^\Gamma$ is given by:

\begin{align}W_{1}= & \frac{1}{4}\Big[ \mathbb{I}\otimes\mathbb{I}+\sigma_{z}\otimes\sigma_{z}+(a^{2}-b^{2})\sigma_{x}\otimes\sigma_{x}\\
& +(a^{2}-b^{2})\sigma_{y}\otimes\sigma_{y}+2ab \left(\sigma_{z}\otimes\mathbb{I}+\mathbb{I}\otimes\sigma_{z}\right) \Big],\nonumber
\end{align}
where the presence of only measurements from the set $\mathcal{M}$ is manifest. The list of the others EWs is provided in the Supplementary Material. These operators generalize two qubit EWs derived in \cite{BNT}. Recall that two qubit states of the form $\rho = \frac 14 ( \oper \otimes \oper + \sum_k   c_k  \sigma_k\otimes \sigma_k )$ span so called {\em  magic simplex} \cite{simplex1,simplex2}. 


In the following we show explicitly the performance of the extremal decomposable witnesses on states that emerge from
 an amplitude-damping channel,
with damping parameter $\gamma\text{\ensuremath{\in\left[0,1\right]}}$, applied on one of the two qubits.
This channel has the form
$\mathcal{E}\left(\rho\right)=A_{0}\rho A_{o}^{\dagger}+A_{1}\rho A_{1}^{\dagger}$,
with $A_{0}=\ket{0}\bra{0}+\sqrt{1-\gamma}\ket{1}\bra{1}$ and $A_{1}=\sqrt{\gamma}\ket{0}\bra1$.
If this channel is applied to the second subsystem
of a two qubit-system, with the Bell state $\ket{\phi^{+}}$ as input, the output state $A=\mathbb{I\otimes\mathcal{E}}\left(\ket{\phi^{+}}\bra{\phi^{+}}\right)$ is entangled for $\gamma<1$.
The witness  $W_{2}$ identifies the state $A$ as entangled for any $\gamma$. Indeed, we have: $\Tr\left[W_{2}A\right]=\frac{1}{4}\left[2\left(-1+2a^{2}\right)\sqrt{1-\gamma}+\gamma-2a\sqrt{1-a^{2}}\gamma\right]$
which for $0\leq\gamma<1$, i.e. all the entangled states, can be
negative for some suitable choice of $a$. For example if $\gamma=0.9$,
then $\Tr\left[W_{2}A\right]<0$ for $0.17<a<\frac{1}{\sqrt{2}}$.
While for $\gamma=0.95$ we have $\Tr\left[W_{2}A\right]<0$ for $0.38<a<\frac{1}{\sqrt{2}}$.
It is interesting to note that in order to detect entangled states for $\gamma$ close
to $1$, which are the least entangled, we have
to consider  witnesses with $a$ different from the extreme points
$\pm1$ and $0$.
This shows that our ability to detect entangled
states increases by considering witnesses $W_i$ characterized by $a,b\neq 0,\pm1$, {in contrast to the use of} only the extremal witnesses on the Bell states, which in this case are {fail.}
\\

Let us now explain how the EWs derived {above} can be operationally used.  Given several copies of an unknown two-qubit quantum state $\rho$, if we want to determine whether $\rho$ is entangled or not we can perform the local measurements given in $\mathcal{M}$ and collect the statistics of their measurement outcomes. Then, from the statistics, we can evaluate ${\rm Tr}(W_k \rho) $ for all $k$. If one of these quantities is below zero, then the state is identified as entangled. A classical optimization of the parameters $a,b$ is required to achieve the best procedure performance, {by computing $\min_{k,a,b}\Tr[W_k(a,b) \rho]$. Such a procedure can be easily impremented in an experimental scenario, e.g. quantum optical implementation such as the one considered in \cite{Exp}, where a similar apparatus was used to demonstrate an efficient test to detect quantum channel capacities \cite{ChiaraMax1}.}

\vspace{.1cm}

\noindent {\em Higher dimensions.}--- {In order to extend the above results}, our goal is to construct entanglement witnesses for bipartite qudit systems by considering the statistics of only few local measurements. In a $d$-dimensional Hilbert space a convenient choice of the basis in the space of linear operators is provided by a set of generalized Gell-Mann matrices $G_{\alpha}\ (\alpha=1,2,\ldots,d^2-1)$ \cite{key55,key-22}.  These are traceless hermitian matrices that can be divided into three groups: $(i)$ diagonal matrices:
\begin{equation}
G^{D}_{l}=\sqrt{\frac{1}{l(l+1)}}\left(\sum_{j=1}^{l}\ket{j}\bra{j}-l\ket{l+1}\bra{l+1}\right),
\end{equation}
for $1\leq l\leq d-1$; $(ii)$ symmetric matrices:
\begin{equation}
G^{S}_{jk}=(\ket{j}\bra{k}+\ket{k}\bra{j})/\sqrt{2}, \ \ 1\leq j<k\leq d\label{Lambda S}
\end{equation}
and $(iii)$ antisymmetric ones:
\begin{equation}
G^{A}_{jk}= (\ket{j}\bra{k}-\ket{k}\bra{j})/i\sqrt{2},\ \  1\leq j<k\leq d.\label{Lambda A}
\end{equation}
In what follows we enumerate the Gell-Mann matrices as follows $G_\alpha =\{G^D_l,G^S_{ij},G^A_{ij}\}$.
The normalization factors guarantiee the following orthogonal relations ${\rm Tr}(G_\alpha G_\beta)= \delta_{\alpha \beta}$ for $\alpha=0,1,\ldots,d^2-1$, where $G_0 = \oper/\sqrt{d}$. Now, any Hermitian operator in $\mathbb{C}^d \otimes \mathbb{C}^d$ may be represented as follows $\,  X = \sum_{\alpha,\beta=0}^{d^2-1} x_{\alpha\beta} G_\alpha \otimes G_\beta$. Assuming normalization ${\rm Tr}X=1$  one has

\begin{align} \label{CIA}
X & = \frac{1}{d^2} \Big\{ \oper \otimes \oper + \sum_{\alpha,\beta=1}^{d^2-1}  \Big( a_\alpha G_\alpha \otimes \oper + b_\alpha \oper \otimes G_\alpha \Big) \nonumber \\  & +  \sum_{\alpha,\beta=1}^{d^2-1} C_{\alpha\beta} G_\alpha \otimes G_\beta \Big\} ,
\end{align}
with real generalized Bloch vectors $a_\alpha$, $b_\alpha$, and correlation matrix $C_{\alpha\beta}$. Hence the analog of (\ref{WWW}) corresponds to $C_{\alpha\beta} = c_\alpha \delta_{\alpha\beta}$, i.e. diagonal correlation matrix. \\
The {\em canonical} example of such witnesses is provided by a flip operator:

\begin{equation}\label{F-d}
  \mathbb{F}_d =  \sum_{\alpha=0}^{d^2-1} G_\alpha \otimes G_\alpha    .
\end{equation}
As is well known, $\mathbb{F}_d$ witnesses entanglement within the whole class of $d \otimes d$ Werner states, which generalize to two qudits the class \eqref{werner}.

Interestingly, it was proved in \cite{SIXIA} that for arbitrary orthogonal matrix $O_{\alpha\beta}$ the following operator

\begin{equation}\label{}
 W = \oper \otimes \oper - \sum_{\alpha,\beta=0}^{d^2-1} O_{\alpha\beta} G_\alpha \otimes G_\beta^{\rm T}
\end{equation}
is block-positive and hence it defines an entanglement witness when $W$ has at least one negative eigenvalue. Clearly, $W^\Gamma = \oper \otimes \oper - \sum_{\alpha,\beta=0}^{d^2-1} O_{\alpha\beta} G_\alpha \otimes G_\beta$ is a witness as well. In particular,
$\oper \otimes \oper - \sum_{\alpha=0}^{d^2-1}  G_\alpha \otimes G_\alpha$ defines a witness operator.

In what follows we consider the similar scenario for arbitrary $d$, that is, we look for EWs {of the form \eqref{CIA} that belong to two classes: the class $\mathcal{C}_0$ made by operators with diagonal correlation matrix $C_{\alpha\beta} = c_\alpha \delta_{\alpha\beta}$; and the class  $\mathcal{C}_1$  with correlation matrix that satisfies the following structure:}

\begin{eqnarray}\label{CIIA}
  && \sum_{\alpha,\beta} C_{\alpha\beta} G_\alpha \otimes G_\beta  = \sum_{k,l=1}^{d-1} D_{kl} G^D_k \otimes G^D_l \nonumber \\ && + \sum_{i< j} \Big( S_{ij} G^S_{ij} \otimes G^S_{ij} + A_{ij} G^A_{ij} \otimes G^A_{ij} \Big) .
\end{eqnarray}
Both classes coincide for $d=2$. A straightforward generalization of Theorem 1 consists in the following:

\begin{thm} The following rank-1 projectors belong to $\mathcal{C}_1$:
  \begin{alignat}{2}
\ket{\varphi_{1}}_{jk}= & a\ket{\phi^{+}}_{jk}+b\ket{\phi^{-}}_{jk} \nonumber \\
 \ket{\varphi_{2}}_{jk}= & a\ket{\psi^{+}}_{jk}+b\ket{\psi^{-}}_{jk}\nonumber\\
\ket{\varphi_{3}}_{jk}= & a\ket{\phi^{+}}_{jk}+b\ket{\psi^{+}}_{jk} \\
 \ket{\varphi_{4}}_{jk}= & a\ket{\phi^{-}}_{jk}+b\ket{\psi^{-}}_{jk}\nonumber\\
\ket{\varphi_{5}}_{jk}= & a\ket{\phi^{+}}_{jk}+ib\ket{\psi^{-}}_{jk} \nonumber\\
 \ket{\varphi_{6}}_{jk}= & a\ket{\phi^{-}}_{jk}+ib\ket{\psi^{+}}_{jk} \nonumber
\end{alignat}
where the real $a$ and $b$ satisfy $a^2+b^2=1$. Vectors $\ket{\phi^{\pm}}_{jk}$ and $\ket{\psi^{\pm}}_{jk}$ represent four Bell states defined by  $\ket{\phi^{\pm}}_{jk} = (|jj\rangle \pm |kk\rangle)/\sqrt{2}$ and $\ket{\psi^{\pm}}_{jk} = (|jk\rangle \pm |jk\rangle)/\sqrt{2}$.
\end{thm}
For the proof see the Supplementary Material. Clearly, the above vectors from $\mathbb{C}^d \otimes \mathbb{C}^d$ have Schmidt rank not greater than two. To go beyond Schmidt rank 2 vectors one has the following:

\begin{pro} A rank-1 projector $P_{\rm MC}$ corresponding to a  maximally correlated state $\ket{\psi}_{\rm MC} =\sum_{i}x_{i}\ket{ii}$ with $x_{i}\in\mathbb{R}$ satisfying $\sum_i x_i^2=1$, belongs to $\mathcal{C}_1$. Moreover, it belongs to $\mathcal{C}_0$ iff it is maximally entangled, that is, $x_{i}=\frac{1}{\sqrt{d}}$.
\end{pro}
Indeed, one has
\begin{align}   \label{Pxx}
P_{\rm MC} &= \sum_{i,j=1}^{d}x_{i}x_{j} |i\rangle \langle j| \otimes |i\rangle \langle j| \nonumber \\
& = \sum_{i}x_{i}^{2} |i\rangle \langle i| \otimes |i\rangle \langle i| \\
 & +\frac{1}{2}\sum_{i<j}x_{i}x_{j}\left( G_{S}^{ij}\otimes G_{S}^{ij} - G_{A}^{ij} \otimes G_{A}^{ij} \right) , \nonumber
\end{align}
which proves that it belongs to $\mathcal{C}_1$. If all $x_i = 1/\sqrt{d}$, then it follows from the fact that $dP^+_d = \mathbb{F}^\Gamma$.

This implies that with projectors (\ref{Pxx}) we can build the following EWs: extremal  decomposable $P^\Gamma$ and $W=D-P$ ,
where $D = \sum_{i,j=1}^{d} d_{ij} |i\rangle \langle i| \otimes |j\rangle \langle j|$ is a diagonal matrix. Note, that if $W$ is an EW then $W^\Gamma= D - P^\Gamma$ is an EW as well. In particular, if $D = d \oper \otimes \oper$, then $W= \lambda\, \mathbb{I}\otimes\mathbb{I}-\ket{\psi}\bra{\psi}$,
is an EW iff  $1> \lambda \geq x_{\star}=\max_{i}\left\{ x_{i}\right\}$. Moreover, such $W$ is always decomposable and it is optimal only if $P$ is maximally entangled. To get non-decomposable EW one needs $D \neq \lambda \oper \otimes \oper$.

\begin{ex} Consider the diagonal part

\begin{equation}\label{}
  D =  \sum_{i=1}^{d} |i\rangle \langle i| \otimes \Big( (d-k) |i\rangle \langle i| + \sum_{j=1}^{k} |i+j\rangle \langle i+j | \Big) ,
\end{equation}
with $k\in \{1,\ldots,d-1\}$. It corresponds to $d_{ii}=d-k$, $d_{i,i+1} = \ldots = d_{i,i+k}=1$ and the remaining $d_{ij}=0$.
For $k=1,\ldots,d-2$ it was proved \cite{E2} that $W = D - dP^+_d$ defined non-decomposable EW and for $k=d-1$ it reproduces the reduction EW $\oper \otimes \oper - d P^+_d$. For $d=3$ and $k=1,2$ it gives the celebrated Choi witness which was proved to be also extremal (see also \cite{Maciek} for another analysis of this witness).
\end{ex}

\begin{ex} Consider the diagonal part

\begin{equation}\label{}
  D =  p_0 \sum_{i=1}^{d} |i\rangle \langle i| \otimes \Big( p_0 |i\rangle \langle i| + p_{i-1} |i-1\rangle \langle i-1 | \Big) ,
\end{equation}
for $p_0,p_1,\ldots,p_d > 0$. It  defines a non-decomposable EW $W=D - dP^+_d$ iff: $p_0 \in [d-2,d-1)$ and

$$   p_1 \ldots p_d \geq (d-1 - p_0)^d . $$
Interestingly,  for $d=3$ it was proved that if $p_0=1$ and $p_1 p_2 p_3=1$, then $W$ is also extremal \cite{E1}.
\end{ex}

\begin{ex} Let $d=3$ and consider
\begin{align}
& D_{\left[abc\right]}  =\sum_{i=1}^{3}  |i\rangle \langle i|  \otimes  \Big(  \left(a+1\right)|i\rangle \langle i| \nonumber \\
& + b |i+1\rangle \langle i+1|  +  c |i+2\rangle \langle i+2 |\Big)  , \nonumber
\end{align}
where we add $\mod2$ and $a,b,c\geq0.$ Then, we consider:
\begin{equation*}
W_{\left[abc\right]}=D_{\left[abc\right]}-dP^{+}_3.
\end{equation*}
The above operator is an EW iff \cite{E2}: $a<2$,  $a+b+c\geq2$, and if $a<1$, then additionally $bc>\left(1-a\right)^{2}$.
Note that if $a=0,b=c=1$, then we recover the reduction witness.
The class $W_{\left[abc\right]}$ contains indecomposable EW and therefore
it can be used to detect bound entangled states. A special class is
given by the following choice: $0<a\leq1$, $a+b+c=2$ and $bc=\left(1-a\right)^{2}$,
indeed it contains only extremal witnesses. Moreover, they are indecomposable
iff $b\neq c$.

\end{ex}

\noindent {\em Conclusions}--- In this Letter we have proposed a method to construct EWs from a limited set of local measurements. Such a method completely characterized the class of EW of the form $W = P ^{\Gamma}$ that can be derived from $\mathcal{M}$ on two-qubit systems. The method relies on an optimization procedure performed by classical means on the measurement results and leads to a possible implementation in various experimental, such as the quantum optical implementation considered in \cite{Exp}. Possible generalisations to higher dimensional systems have been proposed. It would be also very interesting to provide similar analysis for different observables, such as the ones defined in \cite{Ali}, and in the multipartite case. \\

\noindent {\em Acknowledgements}--- : We thank O. G\"uhne for useful discussions and comments. \\ D.C. was supported by the Polish National Science Centre project 2015/19/B/ST1/03095.

\appendix

\section{Extremal entanglement witness for two qubits}

The extremal EWs for two-qubit systems derived from the set of local measurements $\sigma_x \otimes \sigma_x$,
 $\sigma_y \otimes \sigma_y$ and  $\sigma_z \otimes \sigma_z$  are given by the following operators:

\begin{align}W_{1}= & \frac{1}{4}\Big[ \mathbb{I}\otimes\mathbb{I}+\sigma_{z}\otimes\sigma_{z}+(a^{2}-b^{2})\sigma_{x}\otimes\sigma_{x}\\
& +(a^{2}-b^{2})\sigma_{y}\otimes\sigma_{y}+2ab \left(\sigma_{z}\otimes\mathbb{I}+\mathbb{I}\otimes\sigma_{z}\right) \Big],\nonumber
\end{align}

\begin{align}W_{2}= & \frac{1}{4}\Big[ \mathbb{I}\otimes\mathbb{I}-\sigma_{z}\otimes\sigma_{z}+(a^{2}-b^{2}) \sigma_{x}\otimes\sigma_{x}\\
 & -(a^{2}-b^{2})\sigma_{y}\otimes\sigma_{y}+2ab (\sigma_{z}\otimes\mathbb{I}-\mathbb{I}\otimes\sigma_{z})\Big],\nonumber
\end{align}

\begin{align}
W_{3}= & \frac{1}{4}\Big[ \mathbb{I}\otimes\mathbb{I}+\sigma_{x}\otimes\sigma_{x}+(a^{2}-b^{2})\sigma_{z}\otimes\sigma_{z}\\
 & +(a^{2}-b^{2})\sigma_{y}\otimes\sigma_{y}+2ab2 (\sigma_{x}\otimes\mathbb{I}+\mathbb{I}\otimes\sigma_{x}) \Big],\nonumber
\end{align}

\begin{align}
W_{4}= & \frac{1}{4}\Big[ \mathbb{I}\otimes\mathbb{I}-\sigma_{x}\otimes\sigma_{x}+(a^{2}-b^{2})\sigma_{z}\otimes\sigma_{z}\\
 & -(a^{2}-b^{2})\sigma_{y}\otimes\sigma_{y}-2ab (\sigma_{x}\otimes\mathbb{I}-\mathbb{I}\otimes\sigma_{x}) \Big],\nonumber
\end{align}

\begin{align}W_{5}= & \frac{1}{4}\Big[ \mathbb{I}\otimes\mathbb{I}+\sigma_{y}\otimes\sigma_{y}+(a^{2}-b^{2})\sigma_{z}\otimes\sigma_{z}\\
 & +(a^{2}-b^{2})\sigma_{x}\otimes\sigma_{x}+2ab(\sigma_{y}\otimes\mathbb{I}+\mathbb{I}\otimes\sigma_{y})\Big],\nonumber
\end{align}

\begin{align}
W_{6}= & \frac{1}{4}\Big[\mathbb{I}\otimes\mathbb{I}-\sigma_{y}\otimes\sigma_{y}+(a^{2}-b^{2})\sigma_{z}\otimes\sigma_{z}\\
 & -(a^{2}-b^{2})\sigma_{x}\otimes\sigma_{x}-2ab(\sigma_{y}\otimes\mathbb{I}-\mathbb{I}\otimes\sigma_{y})\Big] ,\nonumber
\end{align}

\section{Proof of theorem $1$}

\begin{thm1} There are six 1-parameter families of rank-1 projectors: $|\varphi\rangle\langle \varphi|$ of the form
 \begin{align} \label{WWWS}
W = & \alpha\, \oper \otimes\oper +\sum_{k=x,y,z} \Big(  a_k \oper \otimes \sigma_k
 + b_k \sigma_k\otimes \oper \Big) \nonumber \\  & + \sum_{k=x,y,z} c_k  \sigma_k\otimes \sigma_k  ,
\end{align}
which are given by:

\begin{alignat}{1}
\ket{\varphi_{1}}= a\ket{\phi^{+}}+b\ket{\phi^{-}}; &  \ket{\varphi_{2}}= a\ket{\psi^{+}}+b\ket{\psi^{-}};  \nonumber \\
\ket{\varphi_{3}}= a\ket{\phi^{+}}+b\ket{\psi^{+}}; & \ket{\varphi_{4}}= a\ket{\phi^{-}}+b\ket{\psi^{-}}; \nonumber \\
\ket{\varphi_{5}}= a\ket{\phi^{+}}+ib\ket{\psi^{-}}; & \ket{\varphi_{6}}= a\ket{\phi^{-}}+ib\ket{\psi^{+}}, \label{SS}
\end{alignat}
where $\ket{\phi^{\pm}}$ and $\ket{\psi^{\pm}}$ are the Bell states, and $a,b \in \mathbb{R}$ are such that $a^2+b^2=1$.\\
\end{thm1}

\begin{proof}

Any 2-qubit entanglement witness $W$ can be represented as follows:
\begin{equation}\label{WS}
  W = \sum_{\mu,\nu=0}^3 T_{\mu\nu} \sigma_\mu \otimes \sigma_\nu ,
\end{equation}
with $\sigma_\mu \in \{\oper,\sigma_x,\sigma_y,\sigma_z\}$, and 16 real parameters $T_{\mu\nu}$.
Witnesses of the form \eqref{WWWS} have diagonal correlation tensor $T_{ij} = c_i \delta_{ij}$ $(i,j=1,2,3)$
We now derive the most general projectors of the form $\ket{\varphi}\bra{\varphi}$ that satisfy  the conditions $c_{ij}=c_i\delta_{ij}$.
The state pure state $\varphi$ can be decomposed as:
\begin{equation}
\ket{\varphi}=e^{i\chi_{0}}m\ket{00}+e^{i\chi_{i}}n\ket{01}+e^{i\chi_{2}}q\ket{10}+t\ket{11},\label{General state-1}
\end{equation}
where $m^{2}+n^{2}+q^{2}+t^{2}=1.$ The projector $\ket{\varphi}\bra{\varphi}$
consists in $16$ terms that can be expressed in terms of the single
system Pauli operators by using the following equations:

\begin{equation}
\begin{array}{c}
\ket0\bra0=\frac{1}{2}\left(\mathbb{I}+\sigma_{z}\right);\\
\ket0\bra1=\frac{1}{2}\left(\sigma_{x}+i\sigma_{y}\right);\\
\ket1\bra0=\frac{1}{2}\left(\sigma_{x}-i\sigma_{y}\right);\\
\ket1\bra1=\frac{1}{2}\left(\mathbb{I}-\sigma_{z}\right).
\end{array}
\end{equation}
Then we can gather the coefficients of each operators and arrive at
a projector of the form \eqref{WS} . Finally, we impose the
conditions $c_{ij}=0$, for $i\neq j$ which implies a six equations system that it can be expressed as:
\begin{equation}
\left\{ \begin{array}{c}
mn\cos\left(\chi_{0}-\chi_{1}\right)=qt\cos\chi_{2}\\
mn\sin\left(\chi_{0}-\chi_{1}\right)=qt\sin\chi_{2}\\
mq\cos\left(\chi_{0}-\chi_{2}\right)=nt\cos\chi_{1}\\
mq\sin\left(\chi_{0}-\chi_{2}\right)=nt\sin\chi_{1}\\
nq\sin\left(\chi_{1}-\chi_{2}\right)=0\\
mt\sin\chi_{0}=0
\end{array}.\right.\label{eq system 3-1}
\end{equation}
We remind that another equation is provided by the constraint $m^{2}+n^{2}+q^{2}+t^{2}=1,$
that expresse the state normalization.\\ Let us analyze which are the solutions of (\ref{eq system 3-1}).
First a class of solutions can be derived by the two last equations
by imposing that one of the two parameters that multiply the sine
functions vanishes. In this case the solutions are given by the following
values:
\begin{equation}
\begin{array}{ccc}
m^{2}+t^{2}=1, & n=q=0 & \chi_{0}=0,\pi;\end{array}\label{1 sol-1}
\end{equation}
or
\begin{equation}
\begin{array}{ccc}
n^{2}+q^{2}=1, & m=t=0 & \chi_{2}-\chi_{1}=0,\pi.\end{array}\label{2 sol-1}
\end{equation}
Note that we cannot have solutions like $q^{2}+t^{2}=1$ with $b\neq0,1$,
indeed if we impose $m=n=0$ we would have:
\begin{equation}
\left\{ \begin{array}{c}
qt\cos\chi_{2}=0\\
qt\sin\chi_{2}=0\\
q^{2}+t^{2}=1
\end{array},\right.
\end{equation}
which is verified only in the points $q=0,1$ and $t=1,0$. In terms
of states the solutions (\ref{1 sol-1}) and (\ref{2 sol-1}) are
represented by:
\begin{equation}
\ket{\varphi_{m}}=m\ket{00}\pm t\ket{11}
\end{equation}
and
\begin{equation}
\ket{\varphi_{n}}=n\ket{01}\pm q\ket{10}.
\end{equation}
Note that they contain the four Bell states, namely $\ket{\phi^{\pm}}=\frac{1}{\sqrt{2}}\left(\ket{00}\pm\ket{11}\right)$
and $\ket{\psi^{\pm}}=\frac{1}{\sqrt{2}}\left(\ket{01}\pm\ket{10}\right).$
The remaining solutions can be found by imposing $\sin\left(\chi_{1}-\chi_{2}\right)=0$,
i.e. $\chi_{1}-\chi_{2}=0,\pi,$ in (\ref{eq system 3-1}). Besides
(\ref{1 sol-1}) and (\ref{2 sol-1}), which can be restored, new
solutions are linked to the condition $\sin\left(\chi_{0}\right)=0$,
i.e. $\chi_{0}=0,\pi.$ \\
Let us analyze the solutions for $\chi_{1}=\chi_{2}$ and $\chi_{0}=0;$
the other cases produce similar results. We have only one free parameter
$a$ and the solutions are:

\begin{flalign}
n & =\pm\sqrt{\frac{1-2m^{2}}{2}};q=\pm\sqrt{\frac{1-2m^{2}}{2}};\\
t & =-\frac{2abc}{2a^{2}-1};\chi_{2}=0,\pi;\nonumber
\end{flalign}
and
\begin{flalign}
n & =\pm\sqrt{\frac{1-2m^{2}}{2}};q=\pm\sqrt{\frac{1-2m^{2}}{2}};\nonumber \\
t & =\frac{2abc}{2a^{2}-1};\chi_{2}=\frac{\pi}{2},\frac{3\pi}{2}.
\end{flalign}
In order to understand which states correspond to the above solution
we consider the case $n=q=\sqrt{\frac{1-2m^{2}}{2}}$, $\chi_{2}=\frac{\pi}{2}$
and therefore $t=-m.$ Equation (\ref{General state-1}) becomes:
{\small{}
\begin{flalign}
\ket{\varphi}= & m\ket{00}+i\sqrt{\frac{1-2m^{2}}{2}}\left(\ket{01}+\ket{10}\right)-m\ket{11}\nonumber \\
= & \sqrt{2}m\left(\frac{\ket{00}-\ket{11}}{\sqrt{2}}\right)+i\sqrt{1-2m^{2}}\left(\frac{\ket{01}+\ket{10}}{\sqrt{2}}\right)\label{sol states-1}\\
= & \sqrt{2}m\ket{\phi^{-}}+i\sqrt{1-2m^{2}}\ket{\psi^{+}}\nonumber \\
= & a\ket{\phi^{-}}+ib\ket{\psi^{+}}\nonumber
\end{flalign}
}where $a^{2}+b^{2}=1.$ Similar results can be obtained by considering
the others solutions. Finally we arrive at 6 distinct states, which
in terms of the Bell states are:
\begin{alignat}{2}
\ket{\varphi_{1}}= & a\ket{\phi^{+}}+b\ket{\phi^{-}}; & \ket{\varphi_{2}}= & a\ket{\psi^{+}}+b\ket{\psi^{-}};\label{S1-1-1-2}\\
\ket{\varphi_{3}}= & a\ket{\phi^{+}}+b\ket{\psi^{+}}; & \ket{\varphi_{4}}= & a\ket{\phi^{-}}+b\ket{\psi^{-}};\label{S2-1-1-2}\\
\ket{\varphi_{5}}= & a\ket{\phi^{+}}+ib\ket{\psi^{-}}; & \ket{\varphi_{6}}= & a\ket{\phi^{-}}+ib\ket{\psi^{+}};\label{S6-1-1-2}
\end{alignat}
Note that we must have $a^{2}+b^{2}=1$. Moreover the states $\ket{\varphi_{m}}$
and $\ket{\varphi_{n}}$ can be generated from those above. For example
$\ket{\varphi_{m}}$ can be represented in terms of the state $\ket{\varphi_{1}}$
by redefining the parameters.

\end{proof}

\section{Proof of theorem 2}

Any Hermitian operator $X$ in $\mathbb{C}^d \otimes \mathbb{C}^d$ may be represented as
$\,  X = \sum_{\alpha,\beta=0}^{d^2-1} x_{\alpha\beta} G_\alpha \otimes G_\beta$, where the $G_\alpha =\{G^D_l,G^S_{ij},G^A_{ij}\}$  are the Generalized Gell-Mann (GGM) matrices, which can be divided into three types defined as follows: $(i)$ diagonal matrices:
\begin{equation}
G^{D}_{l}=\sqrt{\frac{1}{l(l+1)}}\left(\sum_{j=1}^{l}\ket{j}\bra{j}-l\ket{l+1}\bra{l+1}\right),
\end{equation}
for $1\leq l\leq d-1$; $(ii)$ symmetric matrices:
\begin{equation}
G^{S}_{jk}=(\ket{j}\bra{k}+\ket{k}\bra{j})/\sqrt{2}, \ \ 1\leq j<k\leq d\label{Lambda S}
\end{equation}
and $(iii)$ antisymmetric ones:
\begin{equation}
G^{A}_{jk}= (\ket{j}\bra{k}-\ket{k}\bra{j})/i\sqrt{2},\ \  1\leq j<k\leq d.\label{Lambda A}
\end{equation}
Assuming normalization ${\rm Tr}X=1$  one has
\begin{align} \label{CIS}
X & = \frac{1}{d^2} \Big\{ \oper \otimes \oper + \sum_{\alpha,\beta=1}^{d^2-1}  \Big( a_\alpha G_\alpha \otimes \oper + b_\alpha \oper \otimes G_\alpha \Big) \nonumber \\  & +  \sum_{\alpha,\beta=1}^{d^2-1} C_{\alpha\beta} G_\alpha \otimes G_\beta \Big\} ,
\end{align}
with real generalized Bloch vectors $a_\alpha$, $b_\alpha$, and correlation matrix $C_{\alpha\beta}$. Hence the analog of (\ref{WWW}) corresponds to $C_{\alpha\beta} = c_\alpha \delta_{\alpha\beta}$, i.e. diagonal correlation matrix. \\

Among the operators \eqref{CIS} we distinguish two classes: first, the class  $\mathcal{C}_0$ made by operators with diagonal $C_{\alpha\beta}$; second, the class  $\mathcal{C}_1$ of operators that satisfy:

\begin{eqnarray}\label{CII}
  && \sum_{\alpha,\beta} C_{\alpha\beta} G_\alpha \otimes G_\beta  = \sum_{k,l=1}^{d-1} D_{kl} G_D^k \otimes G_D^l \nonumber \\ && + \sum_{i< j} \Big( S_{ij} G_S^{ij} \otimes G_S^{ij} + A_{ij} G_A^{ij} \otimes G_A^{ij} \Big) .
\end{eqnarray}

\begin{thm2} The following rank-1 projectors belong to $\mathcal{C}_1$:
  \begin{alignat}{2}
\ket{\varphi_{1}}_{jk}= & a\ket{\phi^{+}}_{jk}+b\ket{\phi^{-}}_{jk} \nonumber \\
 \ket{\varphi_{2}}_{jk}= & a\ket{\psi^{+}}_{jk}+b\ket{\psi^{-}}_{jk}\nonumber\\
\ket{\varphi_{3}}_{jk}= & a\ket{\phi^{+}}_{jk}+b\ket{\psi^{+}}_{jk} \\
 \ket{\varphi_{4}}_{jk}= & a\ket{\phi^{-}}_{jk}+b\ket{\psi^{-}}_{jk}\nonumber\\
\ket{\varphi_{5}}_{jk}= & a\ket{\phi^{+}}_{jk}+ib\ket{\psi^{-}}_{jk} \nonumber\\
 \ket{\varphi_{6}}_{jk}= & a\ket{\phi^{-}}_{jk}+ib\ket{\psi^{+}}_{jk} \nonumber
\end{alignat}
where the real $A$ and $b$ satisfy $a^2+b^2=1$. Vectors $\ket{\phi^{\pm}}_{jk}$ and $\ket{\psi^{\pm}}_{jk}$ represent four Bell stare defined by  $\ket{\phi^{\pm}}_{jk} = (|jj\rangle \pm |kk\rangle)/\sqrt{2}$ and $\ket{\psi^{\pm}}_{jk} = (|jk\rangle \pm |jk\rangle)/\sqrt{2}$.
\end{thm2}

\begin{proof}
Without loss of generality we consider the state $\ket{\varphi_{1}}_{jk}$. The projector on $\ket{\varphi_{1}}_{jk}$
is given by: {\small{}
\begin{align}
\ket{\varphi_{1}}_{jk}\bra{\varphi_{1}}_{jk}= & \frac{(1+2ab)}{2}\ket{jj}\bra{jj}+\frac{(1-2ab)}{2}\ket{kk}\bra{kk}\nonumber \\
 & +\frac{(a^{2}-b^{2})}{2}\left(\ket{jj}\bra{kk}+\ket{kk}\bra{jj}\right).
\end{align}
}The projectors $\ket{jj}\bra{jj}$ and $\ket{kk}\bra{kk}$ contain
only terms related to $G^{D}_{l}\otimes G^{D}_{l'}$, $G^{D}_{l}\otimes\mathbb{I}$
and $\mathbb{I}\otimes G^{D}_{l'},$ hence we can focus only
on the remaining two projectors. From equations (\ref{Lambda S})
and (\ref{Lambda A}) we can see that the single system projector
$\ket{j}\bra{k}$ can be expressed as:
\begin{equation}
\ket{j}\bra{k}=\frac{1}{2}\left(G^{S}_{jk}+iG^{A}_{jk}\right),
\end{equation}
hence $\ket{jj}\bra{kk}$ is given by:
\begin{align}
\ket{jj}\bra{kk}= & \frac{1}{4}\left(G^{S}_{jk}\otimes G^{S}_{jk}+iG^{S}_{jk}\otimes G^{A}_{jk}\right)\label{jjkk}\\
 & \frac{1}{4}\left(iG^{A}_{jk}\otimes G^{S}_{jk}-G^{A}_{jk}\otimes G^{A}_{jk}\right).
\end{align}
Its complex conjugated is thus given by:
\begin{align}
\ket{kk}\bra{jj}= & \frac{1}{4}\left(G^{S}_{jk}\otimes G^{S}_{jk}-iG^{S}_{jk}\otimes G^{A}_{jk}\right),\label{kkjj}\\
 & \frac{1}{4}\left(-iG^{A}_{jk}\otimes G^{S}_{jk}-G^{A}_{jk}\otimes G^{A}_{jk}\right),
\end{align}
hence their sum is:
\begin{equation}
\ket{jj}\bra{kk}+\ket{jj}\bra{kk}=\frac{1}{2}\left(G^{S}_{jk}\otimes G^{S}_{jk}-G^{A}_{jk}\otimes  G^{A}_{jk}\right),\label{Sum}
\end{equation}
which contains only terms of the form \eqref{CII},
and so do the projectors on $\ket{\varphi_{1}}_{jk}$. Note that if
we had considered a state of the form $\ket{\varphi_{1}}_{jk}^{\chi}=a\ket{\phi^{+}}_{jk}+e^{i\chi}b\ket{\phi^{-}}_{jk},$
then its corresponding projector would have been expressed by{\small{}
\begin{flalign}
\ket{\varphi_{1}}_{jk}^{\chi}\bra{\varphi_{1}}_{jk}^{\chi}= & \frac{(1+2ab)}{2}\ket{jj}\bra{jj}+\nonumber \\
+\frac{(1-2ab)}{2}\ket{kk}\bra{kk} & +\frac{(a^{2}-b^{2}+2iab\sin\chi)}{2}\ket{jj}\bra{kk}\nonumber \\
 & +\frac{(a^{2}-b^{2}-2iab\sin\chi)}{2}\ket{jj}\bra{kk},
\end{flalign}
} which fits in the class $\mathcal{C_1}$
if and only if $\chi=0,\pi.$ \\
The same derivation, by using the properties of the GGM matrices,
holds for any of the states $\ket{\varphi_{m}}_{jk}$, $m=0,1,2,3,4,5,6$
and $1\leq j<k\leq d.$

\end{proof}

\end{document}